\documentclass[twocolumn,aps,pra,showpacs]{revtex4}
\usepackage[T1]{fontenc}
\usepackage[latin9]{inputenc}
\usepackage{amsmath}
\usepackage{amssymb}
\usepackage{graphicx}
\usepackage{esint}

\makeatletter
\@ifundefined{textcolor}{}
{%
 \definecolor{BLACK}{gray}{0}
 \definecolor{WHITE}{gray}{1}
 \definecolor{RED}{rgb}{1,0,0}
 \definecolor{GREEN}{rgb}{0,1,0}
 \definecolor{BLUE}{rgb}{0,0,1}
 \definecolor{CYAN}{cmyk}{1,0,0,0}
 \definecolor{MAGENTA}{cmyk}{0,1,0,0}
 \definecolor{YELLOW}{cmyk}{0,0,1,0}
}

\makeatother

\begin{document}

\title{Stoner ferromagnetism of a strongly interacting Fermi gas in the
quasirepulsive regime}

\author{Lianyi He$^{1,2}$, Xia-Ji Liu$^{3}$, Xu-Guang Huang$^{4}$, and
Hui Hu$^{3}$}

\affiliation{$^{1}$Theoretical Division, Los Alamos National Laboratory, Los
Alamos, New Mexico 87545, USA}

\affiliation{$^{2}$Department of Physics and Collaborative Innovation Center
for Quantum Matter, Tsinghua University, Beijing 100084, China}

\affiliation{$^{3}$Centre for Quantum and Optical Science, Swinburne University
of Technology, Melbourne 3122, Australia}

\affiliation{$^{4}$Physics Department and Center for Particle Physics and Field
Theory, Fudan University, Shanghai 200433, China}

\date{\today}
\begin{abstract}
Recent advances in rapidly quenched ultracold atomic Fermi gases near
a Feshbach resonance have brought about a number of interesting problems, in the
context of observing the long-sought Stoner ferromagnetic phase transition.
The possibility of experimentally obtaining a ``quasirepulsive''
regime in the upper branch of the energy spectrum due to the rapid
quench is currently being debated, and the Stoner transition
has mainly been investigated theoretically by using perturbation theory or at high
polarization, due to the limited theoretical approaches in the strongly
repulsive regime. In this work, we present a nonperturbative theoretical
approach to the quasirepulsive upper branch of a Fermi gas near a
broad Feshbach resonance, and we determine the finite-temperature phase
diagram for the Stoner instability. Our results agree well with the
known quantum Monte-Carlo simulations at zero temperature, and we recover
the known virial expansion prediction at high temperature for arbitrary
interaction strengths. At resonance, we find that the Stoner transition
temperature becomes of the order of the Fermi temperature, around which
the molecule formation rate becomes vanishingly small. This suggests
a feasible way to observe Stoner ferromagnetism in the nondegenerate
temperature regime. 
\end{abstract}


\maketitle
%%%%%%%%%%%%%%%%%%%%%%%%%%%%%%%%%%%%%%%%%%%%%%%

\section{Introduction}

\label{s1} %%%%%%%%%%%%%%%%%%%%%%%%%%%%%%%%%%%%%%%%%%%%%%%

An ultracold atomic Fermi gas with tunable contact interactions provides
a paradigm to simulate strongly correlated many-body systems due
to its unprecedented controllability \cite{Bloch2008}. With its strong
attractions, it has paved the way for the crossover from Bardeen-Cooper-Schrieffer
(BCS) superfluidity to Bose-Einstein condensation (BEC) of tightly
bound fermionic pairs \cite{Giorgini2008}, while with its strong repulsions,
it may lead to the confirmation of a text-book result of a ferromagnetic
phase transition predicted nearly a century ago, the so-called Stoner
ferromagnetism \cite{Stoner1933}. However, understanding the nature
of this ferromagnetic transition is still an intriguing and controversial
topic \cite{Massignan2014}. This is largely due to the fact that
the experimental tunability of the repulsive interaction comes at the expense
of severe atom loss \cite{Pekker2011}. The regime of strong
effective repulsion can only be reached by rapidly quenching an attractively
interacting atomic Fermi gas to the meta-stable upper branch of its
energy spectrum near a Feshbach resonance \cite{Pricoupenko2004}.
Initial experimental support of the Stoner ferromagnetic transition was
reported in a strongly interacting Fermi gas of $^{6}$Li atoms \cite{Jo2009}.
However, its existence in the same system was ruled out by more advanced
spin-density fluctuation measurements \cite{Sanner2012}. Recent progress
on repulsive polarons suggests that the Stoner transition may be observable
by using a narrow Feshbach resonance \cite{Kohstall2012} or at low dimensions
\cite{Koschorreck2012}. Triggered by these intriguing experimental
observations, over the past few years, there has been considerable
theoretical interests in Stoner ferromagnetism \cite{Duine2005,LeBlanc2009,Zhai2009,Conduit2009,Conduit2009-2,Dong2010,Liu2010a,Liu2010b,Zhang2010,Li2014,Pilati2010,Chang2011,Heiselberg2011,Saavedra2012,He2012,He2014a,He2014b,Cui2013,Massignan2013}.

Stoner's original idea of a ferromagnetic transition is based on a simple
first-order perturbation theory \cite{Stoner1933}, which at zero
temperature predicts a smooth transition at $k_{F}a_{s}=\pi/2$ for
a spin-population balanced system, where $k_{F}$ is the Fermi wave vector
and $a_{s}$ is the \textit{s}-wave scattering length. The application
of the second-order perturbation theory improves the threshold to
$k_{F}a_{s}\simeq1.054$ \cite{Duine2005}, but the value is still
too large to validate the perturbation theory. Recent zero-temperature
quantum Monte Carlo simulations (QMC) \cite{Conduit2009-2,Pilati2010,Chang2011},
a lowest-order constraint variational calculation \cite{Heiselberg2011},
as well as a non-perturbative ladder approximation calculation \cite{He2012},
suggest instead a transition at $k_{F}a_{s}\simeq0.8-0.9$. On the
other hand, in the limit of large spin imbalance, where the system
may behave like a weakly interacting gas of repulsive polarons, the
ferromagnetic transition could be accurately determined \cite{Massignan2013}.
Yet, a unified theoretical framework, which is valid at all temperatures
and interaction strengths, has yet to be developed.

\begin{figure}
\begin{centering}
\includegraphics[clip,width=0.45\textwidth]{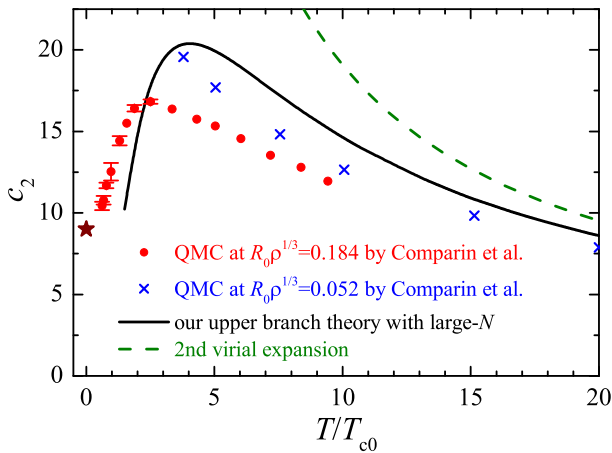} 
\par\end{centering}

\protect\protect\caption{(Color online). The temperature dependence of the two-body contact $c_2$
of a unitary Bose gas. $c_2$ is in units of $\rho^{4/3}$, where $\rho$ is the density of the unitary Bose gas.
$T_{c0}$ is the transition temperature of an ideal Bose gas with density $\rho$. The black solid line shows the prediction of
our upper branch theory within the large-$N$ expansion approximation
\cite{LargeN-Bose}. It agrees reasonably well with the latest Monte
Carlo simulation at both zero temperature (stars, from Ref. \cite{Rossi2014})
and finite temperatures (solid circles and crosses at two different
three-body parameters, from Ref. \cite{QMC-Bose}). The green dashed
line is the second-order virial expansion result that is valid at
high temperatures \cite{LargeN-Bose}. }

\label{fig1} 
\end{figure}

The purpose of this work is to present a nonperturbative theory of
Stoner ferromagnetism at finite temperature, by performing controlled
calculations both in a large-\textit{N} expansion \cite{Nikolic2007,Veillette2007,Enss2012}
and in a dimensional $\epsilon$ expansion \cite{Nishida2006,Nishida2007a,Nishida2007b}.
Previously, the nonperturbative approach with a large-\textit{N} expansion
was applied to study the strongly interacting Bose gas \cite{LargeN-Bose},
which is viewed as the upper branch of a Bose gas across a Feshbach
resonance. Our prediction of Tan's contact density agrees with
the latest results from first-principles quantum Monte Carlo calculations
\cite{Rossi2014,QMC-Bose}, as shown in Fig. \ref{fig1}. In particular,
the nonmonotonic temperature dependence of the two-body contact, predicted
by our theory, is unambiguously confirmed. Therefore, we expect that
the application of this nonperturbative theory to fermions will lead
to a reliable description of the Stoner ferromagnetism at finite temperatures.
A rigorous verification of our predictions can be obtained by confronting
them with more advanced Monte Carlo simulations and experimental investigations
of ferromagnetism at finite temperatures.

%%%%%%%%%%%%%%%%%%%%%%%%%%%%%%%%%%%%%%%%%%%%%%

%%%%%%%%%%%%%%%%%%%%%%%%%%%%%%%%%%%%%%%%%%%%%%%

In this work, we find that the Stoner transition occurs at finite
temperature in a strongly interacting but near-degenerate Fermi gas.
The relatively high transition temperature makes the molecule formation
rate vanishingly small,  and thus the observation of the Stoner transition
will no longer suffer from severe atom loss. Our prediction thereby
paves the way towards experimental confirmation of the long-sought
Stoner ferromagnetic phase transition. Our results may also be used
to better understand the occurrence of ferromagnetism in many strongly
correlated solid-state systems, including superconductors, metals,
and insulators.

One crucial component of our finite-temperature theory is an appropriate
definition of a many-body phase shift for the quasirepulsive upper
branch. In an earlier study, it was realized that a description of
quasirepulsive interaction may be achieved by excluding the in-medium
bound-state contribution from the density equation within a Nozieres-Schmitt-Rink
(NSR) approach \cite{Shenoy2011}. However, this treatment predicts
an equilibrium switch between the upper and the lower branches near
the resonance at high temperature and results in a wide forbidden
area in the low-temperature phase diagram. Alternative spectral representation
of the approach that takes into account an additional frequency-independent
two-body term still suffers from a sudden drop in the spin susceptibility
near the Feshbach resonance \cite{Palestini2012}. Here, we show that
the clarification of the quasirepulsive upper branch, together with
the controllable large-\textit{$N$} expansion and $\epsilon$ expansion,
provides a reliable phase diagram at arbitrary temperatures and coupling
strengths.

Our paper is organized as follows. In the next section (Sec. II),
we review the theoretical framework of the large-$N$ expansion and
the dimensional $\epsilon$ expansion, and we examine the usefulness of
these two approaches for a strongly interacting Fermi gas in the attractive
branch. We explain the definition of the upper branch, Eq. (\ref{eq:phaseshift}),
and we provide a detailed proof of this definition from the viewpoint
of the virial expansion. The technical proof may be skipped for the
first reading. In Sec. III, we present the main results of our work, i.e., 
the Stoner transition at both zero temperature and finite temperature.
A finite-temperature phase diagram is shown and the stability of the
upper branch is briefly discussed. Finally, Sec. IV is devoted to
conclusions and outlooks.

%%%%%%%%%%%%%%%%%%%%%%%%%%%%%%%%%%%%%%%%%%%%%%%

\section{Theory}

\label{s2} %%%%%%%%%%%%%%%%%%%%%%%%%%%%%%%%%%%%%%%%%%%%%%%

We first adopt the large-\textit{N} approach following the pioneering
works by Nikoli\'{c} \textit{et al.} \cite{Nikolic2007} and Veillette
\textit{et al.} \cite{Veillette2007} for an attractive Fermi gas
at the BEC-BCS crossover. An artificial small parameter, $1/N$, is
introduced to organize the different diagrammatic contributions or
scattering processes around the mean-field solution. The original
theory is recovered in the limit of $N=1$. The motivation of the
large-$N$ expansion is that there are no phase transitions with deceasing
$N$, and the large-$N$ (i.e., mean-field) solution has the same symmetry
as the original ground state at $N=1$. Therefore, we anticipate that
the large-$N$ results connect smoothly to the physical results at
$N=1$. One can then perform controlled calculations by including
all diagrams up to a certain order in $1/N$. Although in our calculations
we stop at the next-to-leading-order ($1/N$), systematic improvements
could be achieved by going to higher orders. A complementary approach,
which is similar in spirit, is the dimensional $\epsilon$ expansion.
We will briefly discuss the $\epsilon$ expansion \cite{Nishida2006,Nishida2007a,Nishida2007b}
at the end of this section.

%%%%%%%%%%%%%%%%%%%%%%%%%%%%%%%%%%%%%%%%%%%%%%%

\subsection{Large-\textit{N} expansion}

\label{s2-2} %%%%%%%%%%%%%%%%%%%%%%%%%%%%%%%%%%%%%%%%%%%%%%%

We consider a three-dimensional spin-1/2 interacting Fermi gas with
$N$ fermionic flavors ($i,j=1,\ldots,N$) for each spin degree of
freedom $\sigma=\uparrow,\downarrow$, described by an action (setting
the volume $V=1$ and $\hbar=k_{B}=1$) \cite{Nikolic2007,Veillette2007,Enss2012}
\begin{eqnarray}
\mathcal{S} & = & \int d^{3}\mathbf{x}d\tau\left[\sum_{i=1}^{N}\sum_{\sigma=\uparrow,\downarrow}\psi_{i\sigma}^{*}\left(\partial_{\tau}-\frac{\mathbf{\nabla}^{2}}{2m}-\mu\right)\psi_{i\sigma}\right.\nonumber \\
 &  & \left.+\frac{U_{0}}{N}\sum_{i,j=1}^{N}\psi_{i\uparrow}^{*}\psi_{i\downarrow}^{*}\psi_{j\downarrow}\psi_{j\uparrow}\right],\label{eq:action}
\end{eqnarray}
where $\psi_{i\sigma}\left(\mathbf{x},\tau\right)$ are Grassmann
fields representing fermionic species of equal mass $m$ and the imaginary
time $\tau$ takes values from $0$ to the inverse temperature $\beta=1/T$.
$\mu$ is the chemical potential and $U_{0}$ is the bare interaction
strength to be renormalized in terms of the s-wave scattering length
$a_{s}$ via the relation 
\begin{equation}
\frac{m}{4\pi a_{s}}=\frac{1}{U_{0}}+\sum_{\mathbf{k}}\frac{1}{2\varepsilon_{{\bf k}}}
\end{equation}
with $\varepsilon_{{\bf k}}=\mathbf{k}^{2}/(2m)$. The action possesses
invariance under the symplectic group Sp(2\textit{N}) and in the case
of $N=1$ describes the usual spin-1/2 Fermi gas.

By decoupling the interaction term in the action via a standard Hubbard-Stratonovich
transformation and integrating out the fermionic Grassmann fields,
at the level of Gaussian fluctuations (i.e., the first nontrivial
correction at the order of $1/N$), we obtain the pressure \cite{NSR1985,NSR1993,NSR2006,Liu2006}
\begin{equation}
\frac{\mathcal{P}}{T}=2N\sum_{\mathbf{k}}\ln\left(1+e^{-\beta\xi_{\mathbf{k}}}\right)-\sum_{\mathbf{q},i\nu_{l}}\ln\left[-\Gamma^{-1}\left(\mathbf{q},i\nu_{l}\right)\right],\label{eq:pressure}
\end{equation}
where $\xi_{\mathbf{k}}=\varepsilon_{{\bf k}}-\mu$ and $\Gamma(\mathbf{q},i\nu_{l})$
is the two-particle vertex function with bosonic Matsubara frequencies
$\nu_{l}=2\pi lT$ ($l=0,\pm1,\pm2,\cdots$), 
\begin{equation}
\Gamma^{-1}\left(\mathbf{q},i\nu_{l}\right)=\frac{m}{4\pi a_{s}}-\sum_{\mathbf{k}}\left[\frac{\gamma\left(\mathbf{q},\mathbf{k}\right)}{i\nu_{l}-\xi_{\mathbf{q}/2+\mathbf{k}}-\xi_{\mathbf{q}/2-\mathbf{k}}}+\frac{1}{2\varepsilon_{\mathbf{k}}}\right].\label{eq:inverseVertexFunction}
\end{equation}
$\gamma(\mathbf{q},\mathbf{k})\equiv1-f(\xi_{\mathbf{q}/2+\mathbf{k}})-f(\xi_{\mathbf{q}/2-\mathbf{k}})$
with the Fermi distribution $f(x)=1/(e^{\beta x}+1)$ includes (in-medium)
Pauli blocking of pair fluctuations. By recalling that the vertex
function is essentially the Green function of pairs, the pressure
in Eq. (\ref{eq:pressure}) simply describes a non-interacting mixture
of $2N$ fermionic species and the bosonic pairs \cite{Liu2006}.
By converting the summation over Matsubara frequencies into an integral
over real frequency and introducing an in-medium two-particle phase
shift \cite{NSR1985,NSR1993,NSR2006} 
\begin{equation}
\delta(\mathbf{q},\omega)\equiv-{\rm Imln}[-\Gamma^{-1}(\mathbf{q},\omega+i0^{+})],
\end{equation}
the contribution from the bosonic pairs can be rewritten as 
\begin{equation}
\Delta\mathcal{P}=\sum_{\mathbf{q}}\int_{-\infty}^{+\infty}\frac{d\omega}{\pi}b(\omega)\delta\left(\mathbf{q},\omega\right),\label{eq:pressurePairs}
\end{equation}
where $b(\omega)=1/(e^{\beta\omega}-1)$ is the Bose distribution.
According to standard scattering theory, the two-particle phase shift
is associated with the density of state and increases by $\pi$ when
a two-body bound state emerges \cite{NSR1985}. It should vanish precisely
at $\omega=0$, as required by the integrability of Eq. (\ref{eq:pressurePairs}).

For a unitary Fermi gas in the attractive (ground state) branch, the
application of the large-$N$ expansion has been successful. At zero
temperature, the predicted Bertsch parameter $\xi_{N}=0.279$ \cite{Veillette2007}
is in reasonable agreement with the most recent experimental measurement
$\xi=0.376\pm0.005$ \cite{Ku2012} and quantum Monte Carlo result
$\xi=0.37-0.38$ \cite{Forbes2011,Carlson2011}. The predicted inverse
superfluid transition temperature, $(T_{F}/T_{c})_{N}=6.579$ \cite{Nikolic2007},
is also very close to the experimental data $(T_{F}/T_{c})_{N}=6.0\pm0.5$
\cite{Ku2012}. Near the quantum critical point $\mu=0$, the large-\textit{N}
expansion approach was recently examined by Enss \cite{Enss2012},
by comparing the results for the equation of state and Tan's contact
with more favorable theoretical predictions (i.e., bold diagrammatic
Monte Carlo, BDMC) or the accurate experimental data \cite{Ku2012}.
It was shown that for the pressure $P$, there is excellent agreements
(less than 4\%) between the large-$N$ calculation $P_{N}=0.928nk_{B}T$
and the experimental data $P=0.891\pm0.019nk_{B}T$ (or the BDMC data
$P=0.90\pm0.02nk_{B}T$) \cite{Enss2012}. There is also a similar
good agreement for the entropy density $S/(Nk_{B})$. For Tan's contact
$C$, the large-$N$ prediction ($C_{N}=0.0789k_{F}^{4}$) is just
1.4\% below the BDMC calculation ($C=0.080\pm0.005k_{F}^{4}$). This
is very impressive, given the simplicity of the large-$N$ calculation
\cite{Enss2012}.

%%%%%%%%%%%%%%%%%%%%%%%%%%%%%%%%%%%%%%
\begin{figure}
\begin{centering}
\includegraphics[clip,width=0.45\textwidth]{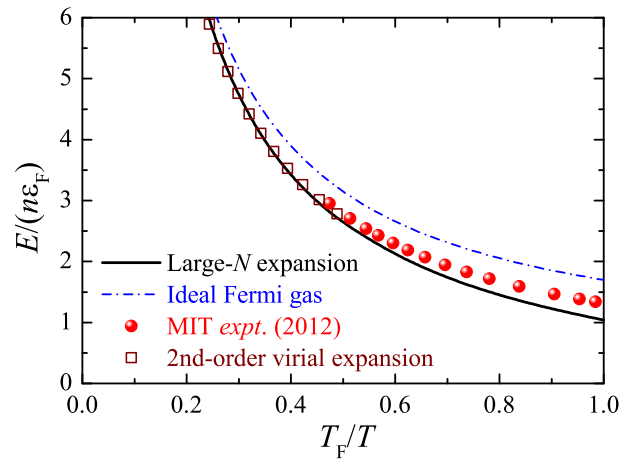} 
\par\end{centering}

\protect\protect\caption{(Color online) Temperature dependence of the total energy of a unitary
Fermi gas predicted by the large-$N$ expansion theory. The large-\textit{N}
expansion results (solid line) are compared with the accurate experimental
measurement from the MIT group (solid circles) \cite{Ku2012}, as
well as the second-order virial expansion prediction (empty squares)
\cite{Liu2013}. The dot-dashed line is the energy of an ideal, non-interacting
Fermi gas.}

\label{fig2} 
\end{figure}

%%%%%%%%%%%%%%%%%%%%%%%%%%%%%%%%%%%%%%%

In Fig. \ref{fig2}, we provide our benchmark of the large-$N$ theory
and systematically compare the large-$N$ predictions with the experimental
results in the non-degenerate regime with $T>T_{F}$. The large-\textit{N}
expansion prediction for the universal energy agrees reasonably well with
the recent experimental measurement by the MIT group \cite{Ku2012}
and with the second-order virial expansion result at sufficiently
large temperatures \cite{Liu2013}.

%%%%%%%%%%%%%%%%%%%%%%%%%%%%%%%%%%%%%%%%%%%%%%%

\subsection{Phase shift of the upper branch}

\label{s2-2} %%%%%%%%%%%%%%%%%%%%%%%%%%%%%%%%%%%%%%%%%%%%%%%

The above theory is only for the attractive branch (ground state).
One crucial component of our finite-temperature theory for the quasirepulsive
upper branch is an appropriate definition of a many-body phase shift.

\begin{figure}
\begin{centering}
\includegraphics[clip,width=0.45\textwidth]{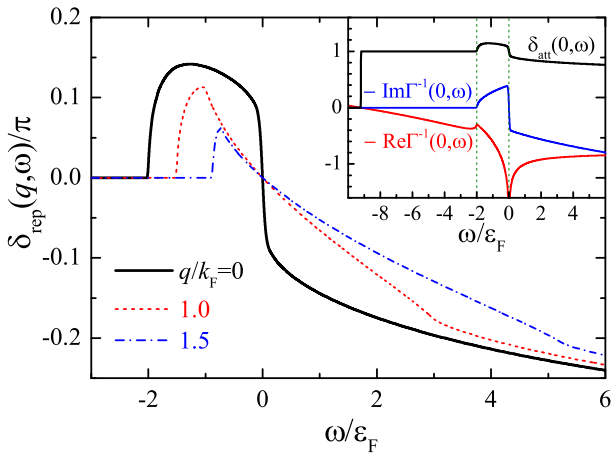} 
\par\end{centering}

\protect\protect\caption{(Color online). The in-medium phase shift $\delta_{\textrm{rep}}(q,\omega)$
of a quasirepulsive Fermi gas at the interaction parameter $k_{F}a_{s}=0.5$
and $T=0$. We have taken a chemical potential $\mu=\varepsilon_{F}$,
which is suitable for the weakly interacting regime. The inset shows
the corresponding in-medium phase shift for an attractive Fermi gas
$\delta_{\textrm{att}}(q=0,\omega)$, together with the real and imaginary
parts of the negative inverse of the two-particle vertex function,
$-\Gamma^{-1}(q=0,\omega)$.}

\label{fig3} 
\end{figure}

For the attractive ground state, the phase shift $\delta_{\textrm{att}}\left(\mathbf{q},\omega\right)$
is generally positive, and the condition $\delta_{\textrm{att}}(\mathbf{q},\omega=0)=0$
is a sufficient criterion to determine the lowest temperature (i.e.,
$T_{c}$) for Cooper pairing instability. In the inset of Fig. \ref{fig3},
we show a typical phase shift for the attractive branch at $1/(k_{F}a_{s})=2$
with a chemical potential at Fermi energy, $\mu=\varepsilon_{F}=k_{F}^{2}/(2m)$.
With increasing frequency, the phase shift jumps from 0 to $\pi$
at a critical value $\omega_{b}(\mathbf{q})$, where the vertex function
develops a pole. This simply signals the existence of a two-body bound
state. With further increasing frequency above the scattering threshold
$\omega_{s}(\mathbf{q})=\mathbf{q}^{2}/(4m)-2\mu>\omega_{b}(\mathbf{q})$,
the phase shift deviates from $\pi$ as the imaginary part of $-\Gamma^{-1}(\mathbf{q},\omega)$
is no longer zero, indicating the scattering continuum. We note that
in this case the criterion $\delta_{\mathbf{\textrm{att}}}(\mathbf{q},\omega=0)=0$
is clearly not satisfied. This is because we have used an unrealistic
large chemical potential for the attractive ground state. In a realistic
solution, the chemical potential will be necessarily pinned by the
Thouless criterion to a value slightly larger than half of the
bound-state energy $-1/(ma_{s}^{2})$ \cite{NSR1985}.

%%%%%%%%%%%%%%%%%%%%%%%%%%%%%%%%%%%%%%%%%%%

%%%%%%%%%%%%%%%%%%%%%%%%%%%%%%%%%%%%%%%%%%%%

For the quasirepulsive upper branch, we first notice that the two-body
phase shift in vacuum is given by 
\begin{equation}
\delta_{2{\rm B}}(E)=-{\rm Im}{\rm ln}\left[-\frac{1}{a_{s}}+\sqrt{-m(E+i0^{+})}\right].
\end{equation}
For the BEC side with $a_{s}>0$, we have 
\begin{eqnarray}
\delta_{2{\rm B}}(E)=\left\{ \begin{array}{r@{\quad,\quad}l}
0 & -\infty<E<\varepsilon_{{\rm B}}\\
\pi & \varepsilon_{{\rm B}}<E<0\\
\pi-\arctan(a_{s}\sqrt{mE}) & E>0,
\end{array}\right.
\end{eqnarray}
where $\varepsilon_{{\rm B}}=-1/(ma_{s}^{2})$ is the bound state
energy level. The $\pi$-jump at $E=\varepsilon_{{\rm B}}$ shows
clearly the existence of a bound state. Therefore, to define a quasirepulsive
two-body system, the $\pi$-shift coming from the bound state should
be subtracted. Inspired by this two-body picture, we find another
repulsive solution for the phase shift, by considering a different
branch cut for the argument of $-\Gamma^{-1}(\mathbf{q},\omega)$,
which differs from $\delta_{\textrm{att}}$ by a constant shift $\pi$
from the scattering threshold: 
\begin{equation}
\delta_{\textrm{rep}}\left(\mathbf{q},\omega\right)=\left[\delta_{\textrm{att}}\left(\mathbf{q},\omega\right)-\pi\right]\Theta\left[\omega-\omega_{s}\left(\mathbf{q}\right)\right].\label{eq:phaseshift}
\end{equation}

In the following, we show that Eq. (\ref{eq:phaseshift}) is an appropriate
prescription of the phase shift for the upper branch from the viewpoint
of virial expansion \cite{Liu2013}. In brief, it is known that the
bosonic contribution $\Delta\mathcal{P}$ contains all two-particle
virial series to infinite order of the fugacity $z=e^{\beta\mu}$
\cite{Liu2013}. It can be expressed as 
\begin{equation}
\Delta\mathcal{P}=\mathcal{P}^{(2)}=\sum_{n=2}^{\infty}\mathcal{P}_{n}^{(2)},
\end{equation}
where $\mathcal{P}_{n}^{(2)}\propto z^{n}$ is the $n$-th two-particle
virial contribution. Since the two-body energy spectrum is known exactly,
we can precisely separate $\mathcal{P}_{n}^{(2)}$ into its contributions
from the bound state and from the scattering continuum. By resumming
only the scattering contributions to all orders in $z$, we obtain
precisely the prescription (\ref{eq:phaseshift}).

At high temperature, the fugacity becomes small, $z=e^{\beta\mu}\ll1$.
The contribution of two-particle scattering process to the pressure
can be expressed as \cite{Liu2013} 
\begin{eqnarray}
{\cal P}^{(2)}=\sum_{n=2}^{\infty}{\cal P}_{n}^{(2)}=\frac{2T}{\lambda_{{\rm dB}}^{3}}\sum_{n=2}^{\infty}b_{n}^{(2)}z^{n},
\end{eqnarray}
where $\lambda_{{\rm dB}}=[2\pi/(mT)]^{1/2}$ is the thermal de Broglie
wavelength and $b_{n}^{(2)}$ is the two-particle contribution to
the $n$-th virial coefficient.

The $n$-order contribution ${\cal P}_{n}^{(2)}$ can be obtained
by making the virial expansion of the pressure $\Delta{\cal P}$.
To this end, we put the dependence on the chemical potential $\mu$
into the distribution functions by using a new variable $E=\omega+2\mu-{\bf q}^{2}/(4m)$.
Then we obtain 
\begin{eqnarray}
{\cal P}^{(2)}=\sum_{{\bf q}}\int_{-\infty}^{\infty}\frac{dE}{\pi}b\left(E+\frac{{\bf q}^{2}}{4m}-2\mu\right)\delta(E,{\bf q}).
\end{eqnarray}
Here the phase shift $\delta(E,{\bf q})$ in terms of $E$ can be
expressed as $\delta(E,{\bf q})=-{\rm Im}{\rm ln}\left[A(E,{\bf q})+iB(E,{\bf q})\right]$,
where the functions $A(E,{\bf q})$ and $B(E,{\bf q})$ are given
by 
\begin{eqnarray}
A(E,{\bf q}) & = & -\frac{1}{a_{s}}+\frac{4\pi}{m}{\rm p. v.}\sum_{{\bf k}}\left[\frac{\gamma({\bf k},{\bf q})}{E-2\varepsilon_{{\bf k}}}+\frac{1}{2\varepsilon_{{\bf k}}}\right],\nonumber \\
B(E,{\bf q}) & = & -\frac{4\pi^{2}}{m}\sum_{{\bf k}}\gamma({\bf k},{\bf q})\delta(E-2\varepsilon_{{\bf k}}).\label{eq:IRAB}
\end{eqnarray}
For $E>0$, p.v. stands for the principal value. The virial expansion
of ${\cal P}^{(2)}$ can be worked out by making use of the expansions
of the Bose and Fermi distribution functions. The distribution functions
can be expanded as 
\begin{eqnarray}
b\left(E+\frac{{\bf q}^{2}}{4m}-2\mu\right)=\sum_{n=1}^{\infty}z^{2n}e^{-n\beta(E+\frac{{\bf q}^{2}}{4m})}
\end{eqnarray}
and 
\begin{eqnarray}
f\left(\varepsilon-\mu\right)=\sum_{n=1}^{\infty}z^{n}(-1)^{n-1}e^{-n\beta\varepsilon}.
\end{eqnarray}
Accordingly, the phase shift $\delta(E,{\bf q})$ can be expanded
as 
\begin{eqnarray}
\delta(E,{\bf q})=\delta_{2{\rm B}}(E)+\sum_{n=1}^{\infty}z^{n}\phi_{n}(E,q),\label{phase}
\end{eqnarray}
where $\delta_{2{\rm B}}(E)$ is the two-body phase shift in the vacuum,

Now we consider the BEC side with $a_{s}>0$. According to the expansion
(\ref{phase}) of the phase shift $\delta(E,{\bf q})$, we can divide
the pressure ${\cal P}^{(2)}$ into four contributions.

(\textbf{A}) The first two contributions come from the leading-order
expansion of the phase shift. Keeping only the vacuum two-body phase
shift $\delta_{2{\rm B}}(E)$, we obtain 
\begin{eqnarray}
{\cal P}_{2{\rm B}}^{(2)}=\sum_{n=1}^{\infty}z^{2n}\sum_{{\bf q}}e^{-\frac{n\beta{\bf q}^{2}}{4m}}\int_{-\infty}^{\infty}\frac{dE}{\pi}e^{-n\beta E}\delta_{2{\rm B}}(E).
\end{eqnarray}
The relative two-body motion and the center-of-mass motion are decoupled
because $\delta_{2{\rm B}}(E)$ depends only on $E$. We can separate
the two-body phase shift into its bound-state part $\delta_{{\rm b}}(E)$
and scattering-state part $\delta_{{\rm s}}(E)$. We have $\delta_{2{\rm B}}(E)=\delta_{{\rm b}}(E)+\delta_{{\rm s}}(E)$,
where 
\begin{eqnarray}
\delta_{{\rm b}}(E) & = & \pi\Theta(E-\varepsilon_{{\rm B}}),\nonumber \\
\delta_{{\rm s}}(E) & = & \left[\delta_{2{\rm B}}(E)-\pi\right]\Theta(E-\varepsilon_{{\rm B}}).\label{separation}
\end{eqnarray}
Accordingly, the pressure ${\cal P}_{2{\rm B}}^{(2)}$ can be divided
into its bound-state contribution ${\cal P}_{{\rm b}}^{(2)}$ and
its scattering-state contribution ${\cal P}_{{\rm s}}^{(2)}$. We
have ${\cal P}_{2{\rm B}}^{(2)}={\cal P}_{{\rm b}}^{(2)}+{\cal P}_{{\rm s}}^{(2)}$,
where 
\begin{eqnarray}
{\cal P}_{{\rm b}}^{(2)} & = & \sum_{n=1}^{\infty}z^{2n}\sum_{{\bf q}}e^{-\frac{n\beta{\bf q}^{2}}{4m}}\int_{-\infty}^{\infty}\frac{dE}{\pi}e^{-n\beta E}\delta_{{\rm b}}(E),\nonumber \\
{\cal P}_{{\rm s}}^{(2)} & = & \sum_{n=1}^{\infty}z^{2n}\sum_{{\bf q}}e^{-\frac{n\beta{\bf q}^{2}}{4m}}\int_{-\infty}^{\infty}\frac{dE}{\pi}e^{-n\beta E}\delta_{{\rm s}}(E).
\end{eqnarray}
Notice that these two contributions are not simply separated by $E<0$
and $E>0$. Completing the integrals over ${\bf q}$ and $E$, we
obtain 
\begin{eqnarray}
{\cal P}_{{\rm b}}^{(2)} & = & \frac{2^{3/2}T}{\lambda_{{\rm dB}}^{3}}\sum_{n=1}^{\infty}\frac{z^{2n}}{n^{5/2}}e^{-n\beta\varepsilon_{{\rm B}}},\nonumber \\
{\cal P}_{{\rm s}}^{(2)} & = & \frac{2^{3/2}T}{\lambda_{{\rm dB}}^{3}}\sum_{n=1}^{\infty}\frac{z^{2n}}{n^{5/2}}\int_{0}^{\infty}\frac{dk}{\pi}e^{-\frac{n\beta k^{2}}{m}}\frac{d\delta(k)}{dk}.
\end{eqnarray}
Here $\delta(k)=-\arctan(ka_{s})$ is now the usual scattering phase
shift without bound state. It becomes evident that ${\cal P}_{{\rm b}}^{(2)}$
and ${\cal P}_{{\rm s}}^{(2)}$ correspond to the bound-state and
scattering-state contributions, respectively. For $n=1$, they recover
the well-known Beth-Uhlenbeck formula of the second virial coefficient.

(\textbf{B}) The other two contributions come from the higher-order
expansions of the phase shift which show explicitly the medium effect.
These contributions can be called medium corrections and can be expressed
as 
\begin{eqnarray}
{\cal P}_{{\rm m}}^{(2)} & = & \sum_{n=1}^{\infty}\sum_{l=1}^{\infty}z^{2n+l}\sum_{{\bf q}}e^{-\frac{n\beta{\bf q}^{2}}{4m}}\nonumber \\
 &  & \times\int_{-\infty}^{\infty}\frac{dE}{\pi}e^{-n\beta E}\phi_{l}(E,q).
\end{eqnarray}
We notice that the relative two-body motion and the center-of-mass
motion cannot be separated for these contributions. To obtain the
expansion coefficients $\phi_{n}(E,q)$, we need to evaluate the expansions
for $A(E,{\bf q})$ and $B(E,{\bf q})$. We thus consider two regimes
of $E$: $E<0$ and $E>0$. We will see that the bound-state and scattering-state
contributions are separated by these two regimes. \\
 (1) $E<0$. In this regime we have $B(E,{\bf q})=0$. The real part
$A(E,{\bf q})$ can be expressed as 
\begin{eqnarray}
A(E,{\bf q})=-\frac{1}{a_{s}}+\sqrt{-mE}+\sum_{n=1}^{\infty}\frac{(-1)^{n-1}}{n}z^{n}{\cal A}_{n}(E,q),
\end{eqnarray}
where the expansion coefficients read 
\begin{eqnarray}
{\cal A}_{n}(E,q)=\frac{16mT}{\pi q}e^{-\frac{n\beta q^{2}}{8m}}\int_{0}^{\infty}pdp\frac{e^{-\frac{n\beta p^{2}}{2m}}\sinh\frac{n\beta pq}{2m}}{p^{2}-mE}.
\end{eqnarray}
In this regime, the medium effect ($z\ll1$) induces a shift of the
bound state pole. Therefore, we have formally 
\begin{eqnarray}
\phi_{l}(E,q)=\pi\sum_{\nu=0}^{l-1}\delta^{(\nu)}(E-\varepsilon_{{\rm B}})\varphi_{l}^{\nu}(\varepsilon_{{\rm B}},q),
\end{eqnarray}
where $\delta^{(\nu)}(x)$ is the $\nu$-th derivative of the Dirac
delta function. The expression of $\varphi_{l}^{\nu}(\varepsilon_{{\rm B}},q)$
is rather complicated and is not shown here. From this formal expression,
the medium correction to the pressure in the region $E<0$ can be
expressed as 
\begin{eqnarray}
{\cal P}_{{\rm mb}}^{(2)} & = & \sum_{n=1}^{\infty}\sum_{l=1}^{\infty}z^{2n+l}\sum_{{\bf q}}e^{-\frac{n\beta{\bf q}^{2}}{4m}}\int_{-\infty}^{0}dEe^{-n\beta E}\nonumber \\
 &  & \times\sum_{\nu=0}^{l-1}\delta^{(\nu)}(E-\varepsilon_{{\rm B}})\varphi_{l}^{\nu}(\varepsilon_{{\rm B}},q).
\end{eqnarray}
The integral over $E$ can be completed and finally ${\cal P}_{{\rm mb}}^{(2)}$
can be formally expressed as 
\begin{eqnarray}
{\cal P}_{{\rm mb}}^{(2)}=\sum_{n=1}^{\infty}\sum_{l=1}^{\infty}z^{2n+l}e^{-n\beta\varepsilon_{{\rm B}}}{\cal H}_{nl}(\varepsilon_{{\rm B}}),
\end{eqnarray}
where ${\cal H}_{nl}(\varepsilon_{{\rm B}})$ is a rather complicated
function of $\varepsilon_{{\rm B}}$ (and also $T$) and will not
be shown here. \\
 (2) $E>0$. In this regime we have 
\begin{eqnarray}
A(E,{\bf q}) & = & -\frac{1}{a_{s}}+\sum_{n=1}^{\infty}\frac{(-1)^{n-1}}{n}z^{n}{\cal A}_{n}(E,q),\nonumber \\
B(E,{\bf q}) & = & -\sqrt{mE}+\sum_{n=1}^{\infty}\frac{(-1)^{n-1}}{n}z^{n}{\cal B}_{n}(E,q),
\end{eqnarray}
where 
\begin{eqnarray}
{\cal A}_{n} & = & \frac{16mT}{\pi q}e^{-\frac{n\beta q^{2}}{8m}}{\rm v.p.}\int_{0}^{\infty}pdp\frac{e^{-\frac{n\beta p^{2}}{2m}}\sinh\frac{n\beta pq}{2m}}{p^{2}-mE},\nonumber \\
{\cal B}_{n} & = & \frac{4mT}{q}e^{-\frac{n\beta q^{2}}{8m}}e^{-\frac{n\beta E}{2}}\sinh\frac{n\beta q\sqrt{mE}}{2m}.
\end{eqnarray}
The expansion coefficients $\phi_{l}(E,q)$ can be worked out but
rather lengthy. Formally, it can be expressed as 
\begin{eqnarray}
\phi_{l}(E,q)=\frac{m}{2k}\sum_{\nu=1}^{l}\frac{d^{\nu}\delta(k)}{dk^{\nu}}\eta_{l}^{\nu}(k,q),
\end{eqnarray}
where we have set $E=k^{2}/m$ and $\delta(k)=-\arctan(ka_{s})$ is
again the usual scattering phase shift without bound state. The function
$\eta_{l}^{\nu}(k,q)$ is also rather lengthy and will not be shown
here. Then the medium correction to the pressure in the scattering
continuum $E>0$ can be expressed as 
\begin{eqnarray}
{\cal P}_{{\rm ms}}^{(2)} & = & \sum_{n=1}^{\infty}\sum_{l=1}^{\infty}z^{2n+l}\sum_{{\bf q}}e^{-\frac{n\beta{\bf q}^{2}}{4m}}\nonumber \\
 &  & \times\int_{0}^{\infty}\frac{dk}{\pi}e^{-\frac{n\beta k^{2}}{m}}\sum_{\nu=1}^{l}\frac{d^{\nu}\delta(k)}{dk^{\nu}}\eta_{l}^{\nu}(k,q).
\end{eqnarray}

From the above discussions, we find that the pure two-body contributions
${\cal P}_{{\rm b}}^{(2)}$ and ${\cal P}_{{\rm s}}^{(2)}$ cannot
be simply distinguished by the scattering threshold ($E<0$ and $E>0$).
They are given by the separation of the phase shift in Eq. (\ref{separation}).
On the other hand, the medium corrections ${\cal P}_{{\rm mb}}^{(2)}$
and ${\cal P}_{{\rm ms}}^{(2)}$ are separated by the scattering threshold.
We therefore identify ${\cal P}_{{\rm b}}^{(2)}$ and ${\cal P}_{{\rm mb}}^{(2)}$
as the contributions from the bound state and ${\cal P}_{{\rm s}}^{(2)}$
and ${\cal P}_{{\rm ms}}^{(2)}$ as the contributions from the scattering
continuum. Summing only the contributions from the scattering continuum,
we obtain the pressure of the quasirepulsive upper branch, 
\begin{eqnarray}
{\cal P}_{{\rm rep}}^{(2)}={\cal P}_{{\rm s}}^{(2)}+{\cal P}_{{\rm ms}}^{(2)}.
\end{eqnarray}
This result can be finally rewritten in a compact form by using the
fact that 
\begin{eqnarray}
{\cal P}_{{\rm s}}^{(2)}=\sum_{{\bf q}}\int_{0}^{\infty}\frac{dE}{\pi}b\left[E+\omega_{s}({\bf q})\right]\left[\delta_{2{\rm B}}(E)-\pi\right]
\end{eqnarray}
and 
\begin{eqnarray}
{\cal P}_{{\rm ms}}^{(2)}=\sum_{{\bf q}}\int_{0}^{\infty}\frac{dE}{\pi}b\left[E+\omega_{s}({\bf q})\right]\left[\delta(E,{\bf q})-\delta_{2{\rm B}}(E)\right],
\end{eqnarray}
where $\omega_{s}({\bf q})={\bf q}^{2}/(4m)-2\mu$ is the scattering
threshold as defined in the text. We finally obtain 
\begin{eqnarray}
{\cal P}_{{\rm rep}}^{(2)}=\sum_{{\bf q}}\int_{0}^{\infty}\frac{dE}{\pi}b\left[E+\omega_{s}({\bf q})\right]\left[\delta(E,{\bf q})-\pi\right].
\end{eqnarray}
Converting to the variable $\omega$, we obtain 
\begin{eqnarray}
{\cal P}_{{\rm rep}}^{(2)}=\sum_{{\bf q}}\int_{\omega_{s}({\bf q)}}^{\infty}\frac{d\omega}{\pi}b\left(\omega\right)\left[\delta({\bf q},\omega)-\pi\right].
\end{eqnarray}
This result can be re-expressed in terms of the phase shift $\delta_{{\rm rep}}$
for the upper branch, 
\begin{eqnarray}
{\cal P}_{{\rm rep}}^{(2)}=\sum_{{\bf q}}\int_{-\infty}^{\infty}\frac{d\omega}{\pi}b\left(\omega\right)\delta_{{\rm rep}}({\bf q},\omega),\label{upperbranch}
\end{eqnarray}
where the phase shift $\delta_{\textrm{rep}}\left(\mathbf{q},\omega\right)$
is given by Eq. (\ref{eq:phaseshift}). Therefore, we have shown that,
by resumming the two-particle virial contributions from the scattering
continuum to all orders in the fugacity $z$, we obtain precisely
the formulation of the quasirepulsive upper branch, Eq. (\ref{upperbranch}).

The above discussions are based on the assumption of a small fugacity
$z\ll1$. However, it is natural to generalize Eq. (\ref{upperbranch})
to the low temperature region since we have resummed the two-body
virial contributions from the scattering continuum to infinite order
in the fugacity $z$.

The resulting phase shift $\delta_{\textrm{rep}}\left(\mathbf{q},\omega\right)$
for the upper branch is shown in Fig. \ref{fig3}. It varies smoothly
as a function of frequency, vanishes identically at $\omega=0$ and
has the correct negative (positive) sign at positive (negative) frequency,
consistent with a phase shift for repulsive interactions. Our prescription
of the upper branch is similar but different from the excluded-molecular-pole
approximation (EMPA) proposed earlier \cite{Shenoy2011}. We can show
that the EMPA adopts a different phase shift (see Appendix A)
for the upper branch which leads to a sudden drop of the interaction
energy near the resonance and hence an equilibrium switch between
the upper and lower branches. With our prescription, one can reach
the repulsive unitary limit. The violation of Tan's adiabatic theorem
near the resonance \cite{Tan2008a,Tan2008b}, as predicted by the
EMPA \cite{Shenoy2011}, can be avoided. Together with the controllable
large-$N$ expansion and $\epsilon$ expansion introduced below, we
are able to access the widely forbidden low-temperature regime, which
was previously found to be mechanically unstable \cite{Shenoy2011}.
Furthermore, by extending the prescription (\ref{eq:phaseshift})
to the BCS side with $a_{s}<0$, we can recover the full upper branch
as first suggested by Pricoupenko and Castin \cite{Pricoupenko2004}.

%%%%%%%%%%%%%%%%%%%%%%%%%%%%%%%%%%%%%%%%%%%
\begin{figure}
\begin{centering}
\includegraphics[clip,width=0.45\textwidth]{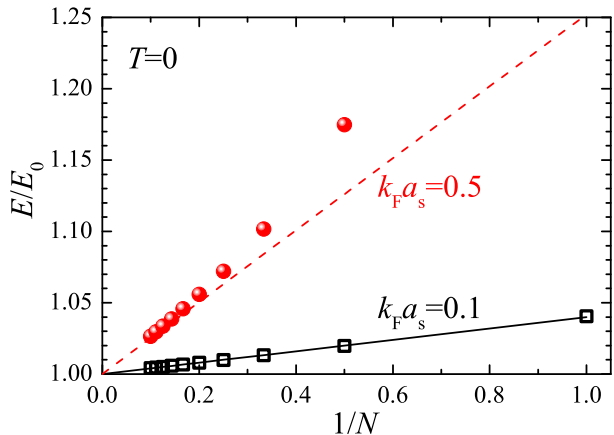} 
\par\end{centering}

\protect\protect\caption{(Color online) The zero-temperature total energy of a quasi-repulsive
Fermi gas. The energy is shown in units of the non-interacting energy
$E_{0}=(3/5)n\varepsilon_{F}$, as a function of $1/N$ at two interaction
parameters: $k_{F}a_{s}=0.1$ (empty squares) and $k_{F}a_{s}=0.5$
(solid circles). The lines are the contribution from the linear part.}

\label{fig4} 
\end{figure}

%%%%%%%%%%%%%%%%%%%%%%%%%%%%%%%%%%%%%%%%%%%%

In Fig. \ref{fig4}, we show the $1/N$-dependence of the energy of an
upper branch Fermi gas at two interaction strengths. At weak interactions
($k_{F}a_{s}=0.1$), the dependence is essentially linear and the
use of the leading $1/N$ term is reasonable. For strong interaction
strengths ($k_{F}a_{s}=0.5$), the dependence is highly non-linear,
due to the unrealistic account of high-order pair fluctuations. In
this case, it is physical to keep only the leading linear term of
the order $1/N$. The higher-order pair fluctuations should be taken
into account by organizing more diagrams beyond Gaussian fluctuations
(i.e., the single bosonic loop) and going to the next-to-next-to-leading
order $O(1/N^{2})$.

%%%%%%%%%%%%%%%%%%%%%%%%%%%%%%%%%%%%%%%%%%

\subsection{Dimensional $\epsilon$ expansion}

\label{s2-3} %%%%%%%%%%%%%%%%%%%%%%%%%%%%%%%%%%%%%%%%%%%

The dimensional $\epsilon$-expansion theory is another non-perturbative
theory developed by Nishida and Son for strongly interacting unitary
Fermi gases \cite{Nishida2006,Nishida2007a,Nishida2007b}. This approach
is based on an expansion around four or two spatial dimensions, where
the pair propagator (or Green function) of Cooper pairs is shown to
be a small quantity \cite{Nishida2007a}. Therefore, one may use the
small number $\epsilon=4-d$ (near four spatial dimensions) or $\bar{\epsilon}=d-2$
(near two spatial dimensions) as a parameter to control the perturbation
expansion. It was found that even at $\epsilon=1$ and $\bar{\epsilon}=1$
the expansion series is reasonably well-behaved, suggesting that it
would be practically useful. Indeed, the next-to-leading-order (NLO)
expansion of a unitary Fermi gas already leads to a surprisingly accurate
Bertsch parameter at zero temperature, $\xi_{\rm{NLO}}=0.377\pm0.014$ \cite{Nishida2009},
which is very close to the most recent experimental result $\xi=0.376\pm0.005$
\cite{Ku2012} and quantum Monte Carlo result $\xi=0.37-0.38$ \cite{Forbes2011,Carlson2011}.
The predicted superfluid transition temperature $(T_{c}/T_{F})_{\rm{NLO}}=0.183\pm0.014$
\cite{Nishida2007b} also agrees very well with the measurement $T_{c}/T_{F}=0.167\pm0.013$
\cite{Ku2012}. For the attractive unitary Fermi gas, one advantage
of the dimensional $\epsilon$ expansion is that the Padé (or Borel-Padé)
approximation can be employed to \emph{match} the expansions around
four and two spatial dimensions and therefore improves the series
summations \cite{Nishida2007a,Nishida2007b,Nishida2009,Arnold2007}.

\begin{figure}
\begin{centering}
\includegraphics[clip,width=0.45\textwidth]{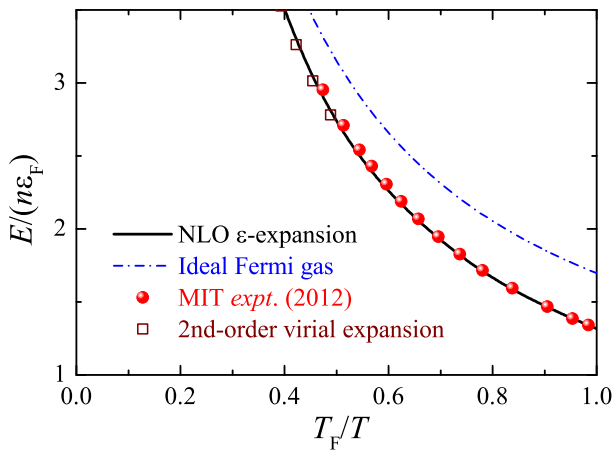} 
\par\end{centering}

\protect\protect\caption{(Color online) Temperature dependence of the total energy of a unitary
Fermi gas predicted by the dimensional $\epsilon$ expansion theory.
As in Fig. \ref{fig3}, the next-to-leading-order (NLO)
$\epsilon$ expansion results (solid line) are contrasted with the
MIT data (solid circles), as well as the second-order virial expansion
(empty squares).}

\label{fig5} 
\end{figure}

However, at finite temperature so far the $\epsilon$ expansion theory
has only been implemented right at the superfluid transition temperature
$T_{c}$. We have re-formulated the $\epsilon$ expansion theory by
using the functional path-integral approach and have made the numerical
calculations practically easy at finite temperature \cite{Mulkerin2016}.
In Fig. \ref{fig5}, we compare the $\epsilon$ expansion results
(extrapolated from $d=4$) for the universal energy $E(T/T_{F})$
of a ground state unitary Fermi gas with the experimental measurement
reported by the MIT group \cite{Ku2012}. The agreement is impressively
good. This is consistent with the excellent agreement found earlier
at zero temperature and at the superfluid transition temperature.
All these agreements strongly suggest that the picture of the unitary
Fermi gas as a mixture of weakly interacting fermionic and bosonic
quasi-particles - which is true near four or two spatial dimensions
- could also be a useful starting point even in three spatial dimensions.

Given the fact that there are no phase transitions by changing the
dimensionality of the system between $d=4$ and $d=2$, we hope the
\emph{predictive} power of the $\epsilon$ expansion theory may persist
as well for a unitary Fermi gas in its quasi-repulsive branch. We
note that in the context of statistical physics, the dimensional $\epsilon$
expansion has been extremely successful and has been applied to describe
the continuous phase transition close to a critical point \cite{Brezin1973,Ma1974,Moshe2003}.

%%%%%%%%%%%%%%%%%%%%%%%%%%%%%%%%%%%%%%%%%

%%%%%%%%%%%%%%%%%%%%%%%%%%%%%%%%%%%%%%%%

%%%%%%%%%%%%%%%%%%%%%%%%%%%%%%%%%%%%%%%%%%%%%%%

\section{Results and discussion}

\label{s3} %%%%%%%%%%%%%%%%%%%%%%%%%%%%%%%%%%%%%%%%%%%%%%%

We have performed numerical calculations for arbitrary coupling strength
$k_{F}a_{s}>0$ and temperature $T$. For strong coupling at low temperature,
the contribution $\Delta\mathcal{P}$ becomes very significant and
highly nonlinear. However, our calculations are still controllable
with the choice of a large $N$ or a dimensionality of space close
to $d=4$. For the large-$N$ expansion, typically, we solve the chemical
potential $\mu$ self-consistently by using the number equation $n=\partial\mathcal{P}/\partial\mu$
for $N=50-100$, where $n=Nk_{F}^{3}/(3\pi^{2})$ is the number density.
Then, we use the large-$N$ expansion $\mu(N)=\mu_{0}+\mu_{1}/N+o(1/N)$
around the non-interacting chemical potential $\mu_{0}$ to extract
the first nontrivial correction $\mu_{1}$ due to pair fluctuations.
The final extrapolation to the $N=1$ limit leads to $\mu=\mu_{0}+\mu_{1}$.
We apply similar expansions to the total energy, inverse compressibility
and inverse spin susceptibility.

\begin{figure}
\begin{centering}
\includegraphics[clip,width=0.45\textwidth]{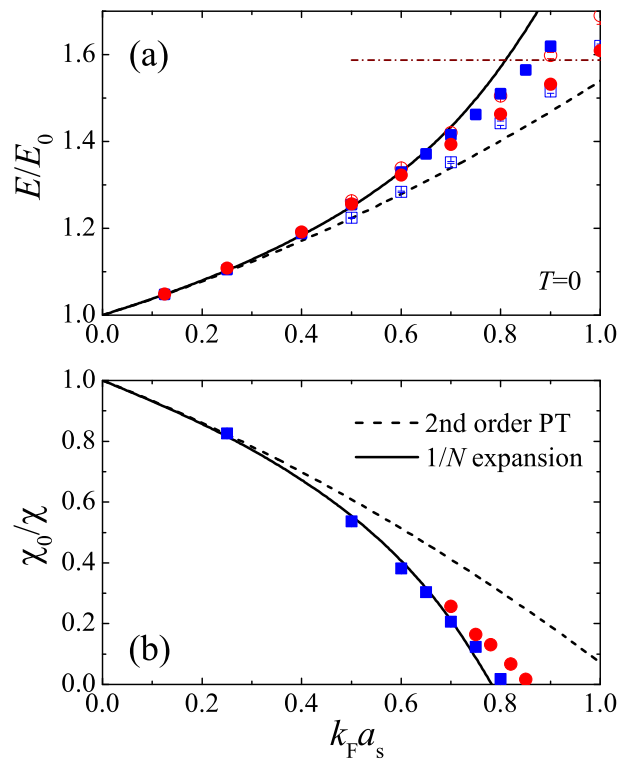} 
\par\end{centering}

\protect\protect\caption{(Color online) The zero-temperature large-$N$ results for the energy
(a) and spin susceptibility (b) of a repulsively interacting Fermi
gas as functions of the interaction parameter $k_{F}a_{s}$, normalized
by the non-interacting results at $T=0$ $E_{0}=(3/5)n\varepsilon_{F}$
and $\chi_{0}=3n/(2\varepsilon_{F})$. For comparison, we also plot
the predictions from the second-order perturbation theory (dashed
line) and quantum Monte-Carlo simulations (symbols). The blue squares
and red circles are the data for the hard-sphere potential and the square-well
potential, respectively. The closed symbols are from Ref. \cite{Pilati2010}
and the open symbols are from Ref. \cite{Chang2011}. The dot-dashed
horizontal line in (a) is the energy of a fully polarized Fermi gas
$E=2^{2/3}E_{0}$.}

\label{fig6} 
\end{figure}

Figure \ref{fig6} reports the interaction parameter dependence of
the energy and inverse spin susceptibility at $T=0$ from the large-$N$
expansion. We find that at weak coupling our large-$N$ expansion
results are consistent with the predictions from second-order
perturbation theory \cite{Galitskii1958}. However, there is an apparent
deviation when the interaction parameter $k_{F}a_{s}>0.4$. It is
impressive that our results agree well with the latest QMC simulations
that use different interaction potentials \cite{Pilati2010,Chang2011}.
In particular, for the inverse spin susceptibility, the agreement
between the large-$N$ expansion and the QMC data for the hard-sphere
potential is exceptionally good. Thus, we determine that at $T=0$
there is a Stoner ferromagnetic transition occurring at $(k_{F}a_{s})_{c}\simeq0.79$,
close to the QMC prediction \cite{Pilati2010}.

%%%%%%%%%%%%%%%%%%%%%%%%%%%%%%%%%%%%%%%%%%%%

%%%%%%%%%%%%%%%%%%%%%%%%%%%%%%%%%%%%%%%%%%%%

%%%%%%%%%%%%%%%%%%%%%%%%%%%%%%%%%%%%%%%%%%%
\begin{figure}
\begin{centering}
\includegraphics[clip,width=0.48\textwidth]{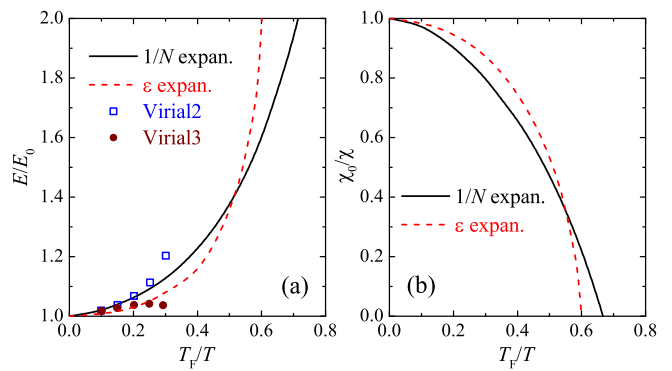} 
\par\end{centering}

\protect\protect\caption{(Color online) The energy (a) and inverse spin susceptibility (b)
of a resonantly interacting Fermi gas in the repulsive regime, normalized
by the corresponding results of an ideal Fermi gas. In (a), we also
show the predictions from the virial expansion theory, up to second
or third order.}

\label{fig7} 
\end{figure}

%%%%%%%%%%%%%%%%%%%%%%%%%%%%%%%%%%%%%%%%%%

Figure \ref{fig7} displays the inverse temperature dependence of
the energy and inverse spin susceptibility of a unitary Fermi gas
in the quasirepulsive branch. At high temperature, our results reproduce
the virial expansion predictions \cite{Liu2010a,Liu2010b,Ho2004,Liu2009,Liu2013},
which are the only known results so far for a repulsive unitary Fermi
gas. It is interesting that, with decreasing temperature down to $(T_{F}/T)_{c}\simeq0.6-0.7$
or $T_{c}\simeq1.5-1.7T_{F}$, both large-$N$ expansion and $\epsilon$
expansion predict a divergent spin susceptibility [see Fig. \ref{fig7}(b)],
signifying the phase transition into a Stoner ferromagnetic state.
The good agreement between the two different non-perturbative theories
strongly indicates that such a transition is realistic and the predicted
transition temperature $T_{c}\simeq1.6T_{F}$ should be qualitatively
reliable.

%%%%%%%%%%%%%%%%%%%%%%%%%%%%%%%%%%%%%%%%%%
\begin{figure}
\begin{centering}
\includegraphics[clip,width=0.45\textwidth]{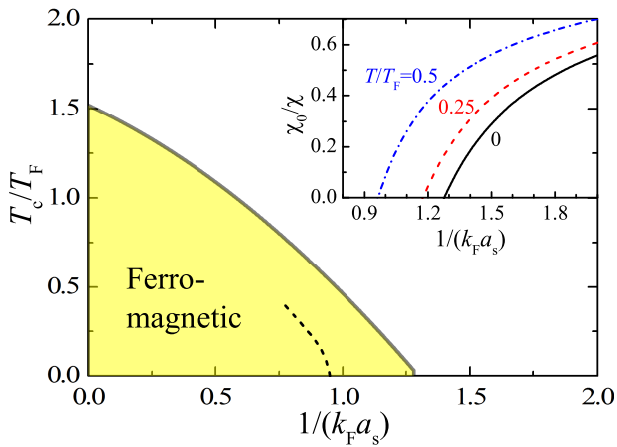} 
\par\end{centering}

\protect\protect\caption{(Color online) Phase diagram of a strongly interacting Fermi gas in
its repulsive regime. In the shadow area, the system energetically
favors spin-domain formation and exhibits Stoner ferromagnetism. The
critical temperature predicted by the second-order perturbation theory
in the low-temperature regime is shown by a dashed line. The inset
shows the inverse spin susceptibility (normalized by the corresponding
non-interacting result) at $T/T_{F}=0$ (solid line), $0.25$ (dashed
line), and $0.5$ (dot-dashed line).}

\label{fig8} 
\end{figure}

%%%%%%%%%%%%%%%%%%%%%%%%%%%%%%%%%%%%%%%%%%

We finally show in Fig. \ref{fig8} a finite temperature phase diagram
of the Stoner ferromagnetism. For comparison, we present also the
prediction from the second order perturbation theory \cite{Duine2005}.
At low temperature, it predicts larger critical interaction parameter;
while close to the unitary limit, it gives unrealistically high transition
temperature due to the strong overestimate of repulsions (not shown
in the figure).

It is worth noting that, in all the cases, including the zero temperature
in Fig. \ref{fig6} or the unitary limit in Fig. \ref{fig7}, the
compressibility of the quasirepulsive Fermi gas predicted by our theory
is always positive. The spin susceptibility is also always well-defined.
Therefore, our approach greatly improves the earlier treatments of
the quasirepulsive upper branch \cite{Shenoy2011,Palestini2012}.

In the experimental studies of Stoner ferromagnetism, the Fermi gas
was originally prepared with weak interactions and then the interactions
were ramped to the strongly repulsive regime. Dynamic rather than
adiabatic preparation was used in order to avoid molecule production.
In the latest experiment of $^{6}$Li Fermi gas at temperature $T\sim0.3T_{F}$,
it was found that the rapid decay into bound pairs (molecules) prevents
the study of equilibrium phases \cite{Sanner2012}. The decay rate
can be theoretically estimated by studying the pair formation rate
$\Delta$, which is given by the imaginary part of the complex pole
of the two-body \textit{T}-matrix \cite{Pekker2011}. At low temperatures ($T<0.5T_{F}$), one
finds a large pair formation rate $\Delta>0.1\varepsilon_{F}$ in
a wide range of the interaction parameter $k_{F}a_{s}$ \cite{Pekker2011},
consistent with the experimental observation of a rapid decay into
bound pairs over times on the order of $10\hbar/\varepsilon_{F}$.

%%%%%%%%%%%%%%%%%%%%%%%%%%%%%%%%%%%%%%%
\begin{figure}
\begin{centering}
\includegraphics[clip,width=0.45\textwidth]{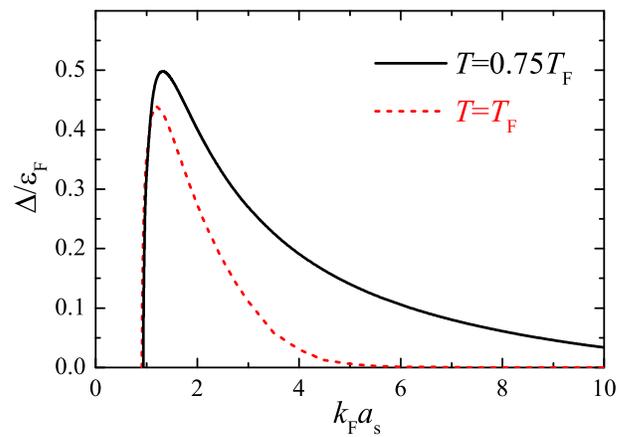} 
\par\end{centering}

\protect\protect\caption{(Color online) The pair formation rate $\Delta$ as a function of
the interaction parameter $k_{F}a_{s}$ at $T=0.75T_{F}$ and $T=T_{F}$. }

\label{fig9} 
\end{figure}

%%%%%%%%%%%%%%%%%%%%%%%%%%%%%%%%%%%%%%

The molecule formation rate can be well estimated by studying the
in-medium two-body \textit{T}-matrix \cite{Pekker2011}. The \textit{T}-matrix
is given by the vertex function $\Gamma({\bf q},\omega)$ but with
the chemical potential $\mu$ replaced by the one for a noninteracting
Fermi gas. The \textit{T}-matrix normally has a complex pole $\omega({\bf q})=\Omega_{{\bf q}}+i\Delta_{{\bf q}}$,
given by the equation $\Gamma^{-1}({\bf q},\omega({\bf q}))=0$. The
imaginary part $\Delta_{{\bf q}}$ characterizes the growth rate of
pair formation in these quenched experiments. For equal spin populations,
the maximal pair formation rate occurs at ${\bf q}=0$. The maximum
pair formation rate $\Delta\equiv\Delta_{{\bf q}=0}$ is determined
by solving the complex pole from the following equation:
\begin{equation}
\frac{1}{a_{s}}-\sqrt{-mE}-\frac{8\pi}{m}\sum_{\mathbf{k}}\frac{f(\xi_{{\bf k}})}{E-2\varepsilon_{\mathbf{k}}}=0.
\end{equation}

In Fig. \ref{fig9}, we examine the pair formation rate at higher
temperatures $T=0.75T_{F}$ and $T=T_{F}$. At large $k_{F}a_{s}$,
the rate is sensitive to the temperature effect. In particular, in the
nondegenerate temperature regime $T\sim T_{F}$, the pair formation
rate becomes vanishingly small for large $k_{F}a_{s}$. In this regime,
it is possible to study the equilibrium phases of strongly repulsive
fermions since the pair formation occurs on a very long time scale
$\gg10\hbar/\varepsilon_{F}$. Therefore, our phase diagram Fig. \ref{fig8}
suggests a promising and realistic way to observe Stoner ferromagnetism
at high temperature and at large interaction parameter $k_{F}a_{s}$.

%%%%%%%%%%%%%%%%%%%%%%%%%%%%%%%%%%%%%%%%%%%%%%%

\section{Summary}

\label{s4} %%%%%%%%%%%%%%%%%%%%%%%%%%%%%%%%%%%%%%%%%%%%%%%

In summary, we have presented a nonperturbative theoretical approach
to the quasirepulsive upper branch of a Fermi gas near a broad Feshbach
resonance, and we determined the finite-temperature phase diagram for the
Stoner instability. One crucial component of our finite-temperature
theory is an appropriate definition of a many-body phase shift for
the quasirepulsive upper branch. We proved this prescription by resumming
the two-body virial contributions from the scattering continuum to
infinite order in the fugacity. Our results agree well with the known
quantum Monte Carlo simulations at zero temperature, and we recover the
known virial expansion prediction at high temperature for arbitrary
interaction strengths. At resonance, the predicted Stoner transition
temperature becomes of order of the Fermi temperature, around which
the molecule formation rate becomes vanishingly small. This suggests
a feasible way to avoid the pairing instability and observe Stoner
ferromagnetism in strongly interacting atomic Fermi gases.

\emph{Note Added.} Recently, we became aware of an experimental work \cite{Roati2016} that reported
the evidence for ferromagnetic instability in the same system as studied
in this paper. In that work,  the detrimental pairing instability was drastically limited
by preparing the gas in a magnetic domain-wall configuration. The
ferromagnetic instability was revealed by observing the softening
of the spin-dipole collective mode that is linked to the increase
of the spin susceptibility. The temperature-coupling phase diagram
was determined \cite{Roati2016}. Our predictions of the critical
gas parameter at $T=0$ [$(k_{F}a)_{c}=0.79$] and the critical temperature
around resonance ($\sim T_{F}$) agrees with their experimental measurements.
\begin{acknowledgments}
We thank Vijay Shenoy and Tin-Lun Ho for useful discussions. L.H.
was supported by the U. S. Department of Energy, Nuclear Physics Office
(Contract No. DE-AC02-05CH11231) and Thousand Young Talent Program
in China. H.H. and X.-J.L. were supported by the ARC Discovery Projects
(Grant Nos. FT130100815, DP140103231, FT140100003 and DP140100637)
and the National Key Basic Research Special Foundation of China (NKBRSFC-China)
(Grant No. 2011CB921502). X.-G.H acknowledges the support from Shanghai
Natural Science Foundation (Grant No. 14ZR1403000) and Fudan University
(Grant No. EZH1512519).
\end{acknowledgments}

\vspace{0.05in}

\appendix
%dummy comment inserted by tex2lyx to ensure that this paragraph is not empty%
%%%%%%%%%%%%%%%%%%%%%%%%%%%%%%%%%%%%%%%%%%%%%%

\section{Comparison with the excluded-molecular-pole approximation}

%%%%%%%%%%%%%%%%%%%%%%%%%%%%%%%%%%%%%%%%%%%%%%%

\begin{figure}
\begin{centering}
\includegraphics[clip,width=0.45\textwidth]{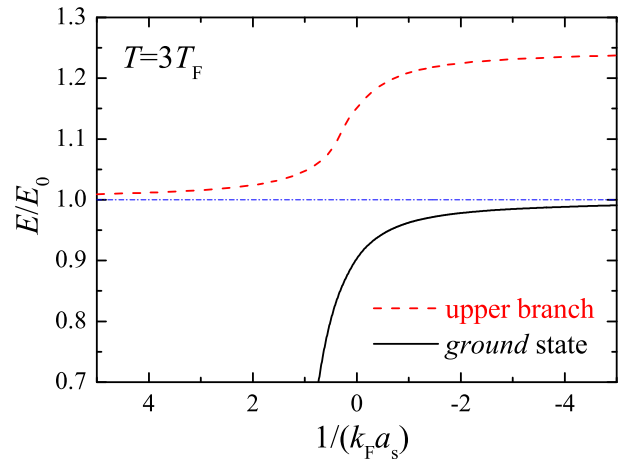} 
\par\end{centering}

\protect\protect\caption{(Color online). The total energy of the upper branch (dashed line)
and of the attractive ground state (solid line) at $T=3T_{F}$, measured
in units of the energy of an ideal, non-interacting Fermi gas.}

\label{fig10} 
\end{figure}

The concept of an upper branch is well-defined for a two-particle system,
where the whole energy spectrum can be solved precisely \cite{Pricoupenko2004,Liu2010a}.
For many-particle systems, however, an unambiguous definition of an upper
branch has yet to be established. Indeed, even for three fermions,
the identification of the upper-branch energy levels turns out to
be difficult \cite{Liu2010a}. To the best of our knowledge, the quasi-repulsive
upper branch of an interacting Fermi gas (at zero temperature) was
first mentioned by Pricoupenko and Castin \cite{Pricoupenko2004},
when they used a lowest-order constraint variational approach to understand
a strongly interacting Fermi gas at the BEC-BCS crossover. The upper
branch prescription provided in this work is a useful extension of
their idea. As a concrete example, in Fig. \ref{fig10} we show the
total energy of the upper branch and the ground-state branch at a
finite temperature $T=3T_{F}$. The generic behavior is not sensitive
to the temperature. The use of the temperature $T=3T_{F}$ makes it 
convenient for us to compare with the result from another approach~\cite{Shenoy2011}.
For the upper branch, in the BEC limit,  the Fermi cloud has a weak
repulsion and its energy approaches the ideal-gas result as $a_{s}\rightarrow0^{+}$.
In the unitary limit with a divergent scattering length, the energy
saturates to a finite value that depends on the temperature. 
%%%%%%%%%%%%%%%%%%%%%%%%%%%%%%%%%%%%%%%%

%%%%%%%%%%%%%%%%%%%%%%%%%%%%%%%%%%%%%%%%

In an earlier work \cite{Shenoy2011}, Shenoy and Ho proposed a different
prescription, the so-called excluded-molecular-pole approximation
(EMPA), for the quasirepulsive upper branch of a strongly interacting
Fermi gas. An important feature of the EMPA is that it predicts an
\emph{equilibrium} branch-switching phenomenon: The energy of the
upper branch reaches a maximum when approaching the resonance from
the BEC side and then changes continuously into the lower branch at
the BCS side. Here we shall compare the EMPA with our approach. 

%%%%%%%%%%%%%%%%%%%%%%%%%%%%%%%%%%%%%%%%
\begin{figure}
\begin{centering}
\includegraphics[clip,width=0.45\textwidth]{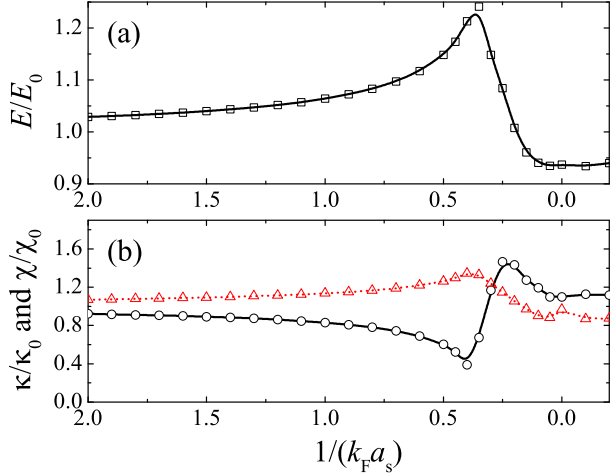} 
\par\end{centering}

\protect\protect\caption{(Color online) The energy, spin susceptibility, and compressibility
as functions of the interaction parameter $1/(k_{F}a_{s})$ at $T=3T_{{\rm F}}$
calculated by starting from the pressure (\ref{PEMPA}) with a simple
phase shift $\delta_{1}({\bf q},\omega)$ given by (\ref{phaseEMPA}).
The results are consistent with those reported in \cite{Shenoy2011}.}

\label{fig11} 
\end{figure}

%%%%%%%%%%%%%%%%%%%%%%%%%%%%%%%%%%%%%%%

A key point of the EMPA is that it starts from the number equation.
In the EMPA, the number density due to two-body interaction is given
by 
\begin{equation}
n_{{\rm rep}}^{(2)}(T,\mu)=\sum_{{\bf q}}\int_{\omega_{s}({\bf q})}^{\infty}\frac{d\omega}{\pi}b(\omega)\frac{\partial\delta({\bf q},\omega)}{\partial\mu},\label{nEMPA}
\end{equation}
where the phase shift is defined as 
\begin{equation}
\delta({\bf q},\omega)=-{\rm Im}{\rm ln}\left[A({\bf q},\omega)+iB({\bf q},\omega)\right].
\end{equation}
The functions $A$ and $B$ are given in Eq. (\ref{eq:IRAB}). We
notice that the ambiguity of the phase shift $\delta({\bf q},\omega)$
is avoided in the above number equation, since it contains only the
derivative of the phase shift with respect to the chemical potential.
In practice, they use the function ${\rm atan}2(y,x)$ \cite{Atan2,Shenoyp}
to evaluate the phase shift and its derivative. Then the two-body
contribution to the pressure can be obtained by using the integration
method, 
\begin{equation}
{\cal P}_{{\rm rep}}^{(2)}(T,\mu)=\int_{-\infty}^{\mu}d\mu^{\prime}n_{{\rm rep}}^{(2)}(T,\mu^{\prime}).
\end{equation}

To compare the EMPA with our prescription, it is convenient to convert
it to an alternative form that starts from the pressure. Toward that
end, we first define a simple phase function 
\begin{equation}
\delta_{1}({\bf q},\omega)=-\arctan\left[\frac{B({\bf q},\omega)}{A({\bf q},\omega)}\right].\label{phaseEMPA}
\end{equation}
Here $\arctan(x)$ is the usual inverse tangent function with a range
$(-\pi/2,\pi/2)$. We have 
\begin{equation}
\frac{\partial\delta({\bf q},\omega)}{\partial\mu}=\frac{\partial\delta_{1}({\bf q},\omega)}{\partial\mu}
\end{equation}
for $\omega>\omega_{s}({\bf q})$. Hence the number equation of the
EMPA can also be expressed as 
\begin{equation}
n_{{\rm rep}}^{(2)}(T,\mu)=\sum_{{\bf q}}\int_{\omega_{s}({\bf q})}^{\infty}\frac{d\omega}{\pi}b(\omega)\frac{\partial\delta_{1}({\bf q},\omega)}{\partial\mu}.\label{n2EMPA}
\end{equation}
The corresponding pressure can be expressed as 
\begin{equation}
{\cal P}_{{\rm rep}}^{(2)}(T,\mu)=\sum_{{\bf q}}\int_{\omega_{s}({\bf q})}^{\infty}\frac{d\omega}{\pi}b(\omega)\delta_{1}({\bf q},\omega).\label{PEMPA}
\end{equation}
The proof of the above result is easy. By taking the derivative of
the above pressure with respect to $\mu$ and using the property of
the phase $\delta_{1}({\bf q},\omega=\omega_{s}({\bf q}))=0$, we
obtain the number equation (\ref{n2EMPA}) and hence (\ref{nEMPA}).
Therefore, the EMPA is equivalent to a scheme starting from the pressure
(\ref{PEMPA}) with a simple phase shift given by (\ref{phaseEMPA}).
This conclusion can also be confirmed numerically. In Fig. \ref{fig11},
we show the results of the energy, spin susceptibility, and compressibility
at $T=3T_{F}$ calculated by starting from the pressure (\ref{PEMPA}).
They are consistent with the results reported in \cite{Shenoy2011}.

In our prescription, the two-body contribution to the pressure is
given by 
\begin{equation}
{\cal P}_{{\rm rep}}^{(2)}(T,\mu)=\sum_{{\bf q}}\int_{\omega_{s}({\bf q})}^{\infty}\frac{d\omega}{\pi}b(\omega)\delta_{{\rm rep}}({\bf q},\omega),
\end{equation}
where 
\begin{equation}
\delta_{\textrm{rep}}\left(\mathbf{q},\omega\right)=\delta_{\textrm{att}}\left(\mathbf{q},\omega\right)-\pi.\label{Ourphase}
\end{equation}
Note that the attractive phase shift $\delta_{\textrm{att}}\left(\mathbf{q},\omega\right)$
should be appropriately determined from its definition $\delta_{{\rm att}}({\bf q},\omega)=-{\rm Im}{\rm ln}\left[A({\bf q},\omega)+iB({\bf q},\omega)\right]$
so that it changes smoothly as a function of $\omega$ for $\omega>\omega_{s}({\bf q})$.
This ensures that the repulsive phase shift $\delta_{{\rm rep}}({\bf q},\omega)$
is also a smooth function of $\omega$ for $\omega>\omega_{s}({\bf q})$
and, in particular, $\delta_{{\rm rep}}({\bf q},\omega=0)=0$ (see
Fig. \ref{fig3} of the text).

In summary, the choice of the phase shift is a nontrivial issue for
a prescription of the upper branch. The EMPA employs the phase shift
$\delta_{1}({\bf q},\omega)$ given by (\ref{phaseEMPA}), while our
prescription adopts the phase shift $\delta_{{\rm rep}}({\bf q},\omega)$
given by (\ref{Ourphase}). Note that our prescription for the phase
shift can be proven by resumming the two-particle virial series to
all orders in the fugacity, as we have shown in the last section.
It is interesting that, in the vacuum, both the phase shifts $\delta_{1}({\bf q},\omega)$
and $\delta_{{\rm rep}}({\bf q},\omega)$ recover the repulsive two-body
phase shift $\delta_{2{\rm B}}(k)=-\arctan(ka_{s})$ (for $a_{s}>0$)
without bound state. On the other hand, we can show that both the
EMPA and our prescription can recover correctly the known perturbative
equation of state at weak coupling $k_{F}a_{s}\rightarrow0^{+}$ and
the second-order virial equation of state in the high temperature
limit. The difference is that, in the EMPA, the upper branch switches
to the lower branch near the resonance. At resonance, the EMPA recovers
the virial equation of state for the lower branch, while our theory
can reach the repulsive unitary limit.

In an early experiment \cite{Bourdel2003}, the interaction energy
of a strongly interacting Fermi gas was measured by using the expansion
properties of a $^{6}$Li gas. At temperature $T\simeq0.6T_{{\rm F}}$,
it was found that the interaction energy of the repulsive branch suddenly
jumps to negative values at magnetic field $B\simeq720$ G, which
lies at the BEC side of the resonance ($k_{F}a_{s}\sim1$). This may
be an experimental support for the EMPA which predicts an \emph{equilibrium}
switch between the upper and the lower branches. However, to our knowledge,
another reasonable explanation for the sudden jump of the interaction
energy at $B\simeq720$ G is the severe \emph{nonequilibrium} atom
loss due to three-body recombination \cite{Sanner2012,Bourdel2003,Massignan2014}.
We believe that, once the atom loss rate can be suppressed by some
effects (such as high temperature, narrow resonance, mass imbalance,
and low dimensionality), one can reach the repulsive unitary limit
experimentally.


\begin{thebibliography}{10}
\bibitem{Bloch2008} I. Bloch , J. Dalibard, and W. Zwerger, \textit{Many-body
physics with ultracold gases}, \emph{Rev. Mod. Phys.} \textbf{80},
885 (2008).

\bibitem{Giorgini2008} S. Giorgini, L. P. Pitaevskii, and S. Stringari,
\emph{Theory of ultracold atomic Fermi gases, Rev. Mod. Phys.} \textbf{80},
1215 (2008).

\bibitem{Stoner1933} E. Stoner, \textit{Atomic moments in ferromagnetic
metals and alloys with nonferromagnetic elements}, \emph{Philos. Mag.}
\textbf{15}, 1018 (1933).

\bibitem{Massignan2014} P. Massignan, M. Zaccanti, G. M. Bruun, \textit{Polarons,
dressed molecules, and itinerant ferromagnetism in ultracold Fermi
gases}, \emph{Rep. Prog. Phys.} \textbf{77}, 034401 (2014).

\bibitem{Pekker2011} D. Pekker, M. Babadi, R. Sensarma, N. Zinner,
L. Pollet, M. W. Zwierlein, and E. Demler, \textit{Competition between
pairing and ferromagnetic instabilities in ultracold Fermi gases near
Feshbach resonances}, \emph{Phys. Rev. Lett.} \textbf{106}, 050402
(2011).

\bibitem{Pricoupenko2004} L. Pricoupenko and Y. Castin, \textit{One
particle in a box: The simplest model for a Fermi gas in the unitary
limit}, \emph{Phys. Rev. A} \textbf{69}, 051601(R) (2004).

\bibitem{Jo2009} G.-B. Jo, Y.-R. Lee, J.-H. Choi, C. A. Christensen,
T. H. Kim, J. H. Thywissen, D. E. Pritchard, and W. Ketterle, \textit{Itinerant
ferromagnetism in a Fermi gas of ultracold atoms}, \emph{Science}
\textbf{325}, 1521 (2009).

\bibitem{Sanner2012} C. Sanner, E. J. Su, W. Huang, A. Keshet, J.
Gillen, and W. Ketterle, \textit{Correlations and pair formation in
a repulsively interacting Fermi gas}, \emph{Phys. Rev. Lett.} \textbf{108},
240404 (2012).

\bibitem{Kohstall2012} C. Kohstall, M. Zaccanti, M. Jag, A. Trenkwalder,
P. Massignan, G. M. Bruun, F. Schreck, and R. Grimm, \textit{Metastability
and coherence of repulsive polarons in a strongly interacting Fermi
mixture},\emph{ Nature (London)} \textbf{485}, 615 (2012).

\bibitem{Koschorreck2012} M. Koschorreck, D. Pertot, E. Vogt, B.
Fröhlich, M. Feld, and M. Köhl, \textit{Attractive and repulsive Fermi
polarons in two dimensions}, \emph{Nature (London)} \textbf{485},
619 (2012).

\bibitem{Duine2005} Duine R. A. \& MacDonald A. H. Itinerant ferromagnetism
in an ultracold atom Fermi gas \emph{Phys. Rev. Lett.} \textbf{95},
230403 (2005).

\bibitem{LeBlanc2009} LeBlanc L. J., Thywissen J. H., Burkov A. A.
\& Paramekanti A. Repulsive Fermi gas in a harmonic trap: Ferromagnetism
and spin textures. \emph{Phys. Rev. A} \textbf{80}, 013607 (2009).

\bibitem{Zhai2009} H. Zhai, \textit{Correlated versus ferromagnetic
state in repulsively interacting two-component Fermi gases}, \emph{Phys.
Rev. A} \textbf{80}, 051605(R) (2009).

\bibitem{Conduit2009} G. J. Conduit and B. D. Simons, \textit{Repulsive
atomic gas in a harmonic trap on the border of itinerant ferromagnetism},
\emph{Phys. Rev. Lett.} \textbf{103}, 200403 (2009).

\bibitem{Conduit2009-2} G. J. Conduit, A. G. Green, and B. D. Simons,
\textit{Inhomogeneous phase formation on the border of itinerant ferromagnetism},
\emph{Phys. Rev. Lett.} \textbf{103}, 207201 (2009).

\bibitem{Dong2010} H. Dong, H. Hu, X.-J. Liu, and P. D. Drummond,
\textit{Mean-field study of itinerant ferromagnetism in trapped ultracold
Fermi gases: Beyond the local-density approximation}, \emph{Phys.
Rev. A} \textbf{82}, 013627 (2010).

\bibitem{Liu2010a} X.-J. Liu, H. Hu, and P. D. Drummond, \textit{Three
attractively interacting fermions in a harmonic trap: Exact solution,
ferromagnetism, and high-temperature thermodynamics}, \emph{Phys.
Rev. A} \textbf{82}, 023619 (2010).

\bibitem{Liu2010b} X.-J. Liu and H. Hu, \textit{Virial expansion
for a strongly correlated Fermi gas with imbalanced spin populations},
\emph{Phys. Rev. A} \textbf{82}, 043626 (2010).

\bibitem{Zhang2010} S. Zhang, H. H. Hung, and C. Wu, \textit{Proposed
realization of itinerant ferromagnetism in optical lattices}, \emph{Phys.
Rev. A} \textbf{82}, 053618 (2010).

\bibitem{Li2014} Y. Li, E. H. Lieb, and C. Wu, \textit{Exact results
on itinerant ferromagnetism in multi-orbital systems on square and
cubic lattices}, \emph{Phys. Rev. Lett.} \textbf{112}, 217201 (2014).

\bibitem{Pilati2010} S. Pilati, G. Bertaina, S. Giorgini, and M.
Troyer, \textit{Itinerant ferromagnetism of a repulsive atomic Fermi
gas: A quantum Monte Carlo study}, \emph{Phys. Rev. Lett.} \textbf{105},
030405 (2010).

\bibitem{Chang2011} S. Y. Chang, M. Randeria, and N. Trivedi, \textit{Ferromagnetism
in repulsive Fermi gases: Upper branch of Feshbach resonance versus
hard spheres}, \emph{Proc. Natl. Acad. Sci. (USA)} \textbf{108},
51 (2011).

\bibitem{Heiselberg2011} H. Heiselberg, \textit{Itinerant ferromagnetism
in ultracold Fermi gases}, \emph{Phys. Rev. A} \textbf{83}, 053635
(2011).

\bibitem{He2012} L. He and X.-G. Huang, \textit{Nonperturbative effects
on the ferromagnetic transition in repulsive Fermi gases}, \emph{Phys.
Rev. A} \textbf{85}, 043624 (2012).

\bibitem{He2014a} L. He, \textit{Finite range and upper branch effects
on itinerant ferromagnetism in repulsive Fermi gases: Bethe-Goldstone
ladder resummation approach}, Ann. Phys. (NY) \textbf{351}, 477
(2014).

\bibitem{He2014b} L. He, \textit{Interaction energy and itinerant
ferromagnetism in a strongly interacting Fermi gas in the absence
of molecule formation}, Phys. Rev. A\textbf{90}, 053633 (2014).

\bibitem{Saavedra2012} F. Arias de Saavedra, F. Mazzanti, J. Boronat,
and A. Polls, \textit{Ferromagnetic transition of a two-component
Fermi gas of hard spheres}, \emph{Phys. Rev. A} \textbf{85}, 033615
(2012).

\bibitem{Cui2013} X. Cui and T.-L. Ho, \textit{Phase separation in
mixtures of repulsive Fermi gases driven by mass difference}, \emph{Phys.
Rev. Lett.} \textbf{110}, 165302 (2013).

\bibitem{Massignan2013} P. Massignan, Z. Yu, and G. M. Bruun, \textit{Itinerant
ferromagnetism in a polarized two-component Fermi gas}, \emph{Phys.
Rev. Lett.} \textbf{110}, 230401 (2013).

\bibitem{Nikolic2007} P. Nikoli\'{c} and S. Sachdev, \textit{Renormalization-group
fixed points, universal phase diagram, and $1/N$ expansion for quantum
liquids with interactions near the unitarity limit}, \emph{Phys. Rev.
A} \textbf{75}, 033608 (2007).

\bibitem{Veillette2007} M. Y. Veillette, D. E. Sheehy, and L. Radzihovsky,
\textit{Large-$N$ expansion for unitary superfluid Fermi gases},
\emph{Phys. Rev. A} \textbf{75}, 043614 (2007).

\bibitem{Enss2012} T. Enss, \textit{Quantum critical transport in
the unitary Fermi gas}, \emph{Phys. Rev. A} \textbf{86}, 013616 (2012).

\bibitem{Nishida2006}Y. Nishida and D. T. Son, \textit{$\epsilon$
expansion for a Fermi gas at infinite scattering length}, Phys. Rev.
Lett. \textbf{97}, 050403 (2006).

\bibitem{Nishida2007a} Y. Nishida and D. T. Son, \textit{Fermi gas
near unitarity around four and two spatial dimensions} Phys. Rev.
A \textbf{75}, 063617 (2007).

\bibitem{Nishida2007b}Y. Nishida, \textit{Unitary Fermi gas at finite
temperature in the $\epsilon$ expansion}, Phys. Rev. A \textbf{75},
063618 (2007).

\bibitem{LargeN-Bose} X.-J. Liu, B. Mulkerin, L. He, and H. Hu, \textit{Equation
of state and contact of a strongly interacting Bose gas in the normal
state}, Phys. Rev. A \textbf{91}, 043631 (2015).

\bibitem{Rossi2014}M. Rossi, L. Salasnich, F. Ancilotto, and F. Toigo,
\textit{Monte Carlo simulations of the unitary Bose gas}, Phys. Rev.
A \textbf{89}, 041602(R) (2014).

\bibitem{QMC-Bose} T. Comparin and W. Krauth, \textit{Momentum distribution
in the unitary Bose gas from first principles}, arXiv: 1604.08870
(2016).

\bibitem{Shenoy2011} V. B. Shenoy and T.-L. Ho, \textit{Nature and
properties of a repulsive Fermi gas in the upper branch of the energy
spectrum}, \emph{Phys. Rev. Lett.} \textbf{107}, 210401 (2011).

\bibitem{Palestini2012} F. Palestini, P. Pieri, and G. C. Strinati,
\textit{Density and spin response of a strongly interacting Fermi
gas in the attractive and quasirepulsive regime}, \emph{Phys. Rev.
Lett.} \textbf{108}, 080401 (2012).

\bibitem{NSR1985} P. Nozieres and S. Schmitt-Rink, \textit{Bose condensation
in an attractive fermion gas: From weak to strong coupling superconductivity},
\emph{J. Low Temp. Phys.} \textbf{59}, 195 (1985).

\bibitem{NSR1993} C. A. R. Sade Melo, M. Randeria, and J. R. Engelbrecht,
\textit{Crossover from BCS to Bose superconductivity: Transition temperature
and time-dependent Ginzburg-Landau theory}, \emph{Phys. Rev. Lett.}
\textbf{71}, 3202 (1993).

\bibitem{NSR2006} H. Hu, X.-J. Liu, and P. D. Drummond, \textit{Equation
of state of a superfluid Fermi gas in the BCS-BEC crossover}, \emph{Europhys.
Lett.} \textbf{74}, 574 (2006).

\bibitem{Liu2006} X.-J. Liu and H. Hu, \textit{BCS-BEC crossover
in an asymmetric two-component Fermi gas}, \emph{Europhys. Lett.}
\textbf{75}, 364 (2006).

\bibitem{Ku2012} M. J. H. Ku, A. T. Sommer, L. W. Cheuk, and M. W.
Zwierlein, \textit{Revealing the superfluid lambda transition in the
universal thermodynamics of a unitary Fermi gas}, \emph{Science} \textbf{335},
563 (2012).

\bibitem{Forbes2011} M. M. Forbes, S. Gandolfi, and A. Gezerlis,
\textit{Resonantly interacting fermions in a box}, Phys. Rev. Lett.
\textbf{106}, 235303 (2011). 

\bibitem{Carlson2011} J. Carlson, S. Gandolfi, K. E. Schmidt, and
S. Zhang, \textit{Auxiliary-field quantum Monte Carlo method for strongly
paired fermions}, Phys. Rev. A\textbf{84}, 061602(R) (2011).

\bibitem{Liu2013} X.-J. Liu, \textit{Virial expansion for a strongly
correlated Fermi system and its application to ultracold atomic Fermi
gases}, \emph{Phys. Rep.} \textbf{524}, 37 (2013).

\bibitem{Tan2008a}S. Tan, \textit{Energetics of a strongly correlated
Fermi gas}, \textit{Ann. Phys. (NY)} \textbf{323}, 2952 (2008).

\bibitem{Tan2008b}S. Tan, \textit{Large momentum part of a strongly
correlated Fermi gas}, \textit{Ann. Phys. (NY)} \textbf{323}, 2971 (2008).

\bibitem{Nishida2009}Y. Nishida, \textit{Ground-state energy of the
unitary Fermi gas from the $\epsilon$ expansion}, Phys. Rev. A \textbf{79},
013627 (2009).

\bibitem{Arnold2007}P. Arnold, J. E. Drut, and D. T. Son, \textit{Next-to-next-to-leading-order
$\epsilon$ expansion for a Fermi gas at infinite scattering length},
Phys. Rev. A \textbf{75}, 043605 (2007).

\bibitem{Mulkerin2016} B. C. Mulkerin, X.-J. Liu, and H. Hu, \textit{Beyond
Gaussian pair fluctuation theory for strongly interacting Fermi gases},
arXiv: 1602.07391 (2016).

\bibitem{Brezin1973}E. Brezin and D. J. Wallace, \textit{Critical
behavior of a classical Heisenberg ferromagnet with many degrees of
freedom}, Phys. Rev. B \textbf{7}, 1967 (1973).

\bibitem{Ma1974}S. K. Ma, \textit{Scaling variables and dimensions},
Phys. Rev. A \textbf{10}, 1818 (1974).

\bibitem{Moshe2003}M. Moshe and J. Zinn-Justin, \textit{Quantum field
theory in the large N limit: A review}, Phys. Rep. \textbf{385}, 69
(2003).

\bibitem{Galitskii1958} V. M. Galitskii, \textit{The energy spectrum
of a non-ideal Fermi gas}, \emph{Sov. Phys. JETP} \textbf{7}, 104
(1958).

\bibitem{Ho2004} T.-L. Ho and E. J. Mueller, \textit{High temperature
expansion applied to fermions near Feshbach resonance}, \emph{Phys.
Rev. Lett.} \textbf{92}, 160404 (2004).

\bibitem{Liu2009} X.-J. Liu, H. Hu, and P. D. Drummond, \textit{Virial
expansion for a strongly correlated Fermi gas}, \emph{Phys. Rev. Lett.}
\textbf{102}, 160401 (2009).


\bibitem{Atan2} For the definition of the function ${\rm atan}2(y,x)$,
see http://en.wikipedia.org/wiki/Atan2.

\bibitem{Shenoyp} V. B. Shenoy and T.-L. Ho (private communications).

\bibitem{Bourdel2003} T. Bourdel, J. Cubizolles, L. Khaykovich, K.
M. F. Magalhaes, S. J. J. M. F. Kokkelmans, G. V. Shlyapnikov, and
C. Salomon, \textit{Measurement of the interaction energy near a Feshbach
resonance in a $^{6}$Li Fermi gas} Phys. Rev. Lett. \textbf{91},
020402 (2003).


\bibitem{Roati2016} G. Valtolina, F. Scazza, A. Amico, A. Burchianti,
A. Recati, T. Enss, M. Inguscio, M. Zaccanti, and G. Roati, \textit{Evidence
for ferromagnetic instability in a repulsive Fermi gas of ultracold
atoms}, arXiv: 1605.07850 (2016).\end{thebibliography}
\end{document}